\documentclass[12pt]{article}
\usepackage{a4,epsf}
\newcommand{\bea}{\begin{eqnarray}}
\newcommand{\eea}{\end{eqnarray}}
\newcommand{\simgt}{\hbox{ \raise3pt\hbox to 0pt{$>$}\raise-3pt\hbox{$\sim$} }}
\newcommand{\simlt}{\hbox{ \raise3pt\hbox to 0pt{$<$}\raise-3pt\hbox{$\sim$} }}

\begin{document}
\begin{titlepage}
\title{
\vspace{2cm}
Neutrino Masses and Bimaximal Mixing\thanks{Work 
supported in part by KBN grants 2P03B08414 and 
2P03B14715, and by BMBF grant POL-239-96.}
}
\author{M.~Je\.zabek$^{a,b)}$
and
Y.~Sumino$^{c)}$
\\ \\ \\ \small
$a)$ Institute of Nuclear Physics,
Kawiory 26a, PL-30055 Cracow, Poland
\\   \small
  $b)$ Department of Field Theory and Particle Physics, University of 
      Silesia, \\[-5pt]  \small
     Uniwersytecka 4, PL-40007 Katowice, Poland\\ 
\small
$c)$ Department of Physics, Tohoku University,
Sendai 980-77, Japan
}
\date{}
\maketitle
\thispagestyle{empty}
\vspace{-4.5truein}
\begin{flushright}
{\bf TP-USl/99/03}\\
{\bf TU-563}\\
{\bf hep-ph/9904382}\\
{\bf April 1999}
\end{flushright}
\vspace{4.0truein}
\begin{abstract}
Solar and atmospheric neutrino anomalies are described 
by bimaximal mixing of three Majorana neutrinos. 
Neutrino oscillations in appearance 
$\nu_e\rightleftharpoons\nu_\mu,\nu_\tau$
and in disappearance $\nu_e$ long baseline and atmospheric
experiments are sensitive to deviations from the
ideal bimaximal mixing. It is suggested that these
deviations may be dominated by a rotation in the
electron--muon plane of the generation space.
Simple seesaw models are described supporting this idea.
The rotation angle is estimated from Fritzsch's relation.
Predictions are presented for oscillations in long baseline 
experiments, for solar neutrinos, and for the rates of 
neutrinoless double beta decays.
\end{abstract}
\end{titlepage}
\section{Introduction}
Neutrino masses and oscillations\cite{Pontec,MNS}
belong to the most exciting area of particle physics. 
It is exactly this area where the Standard Model of 
quarks and leptons is challenged in a most successful 
way. The recent discovery of muon neutrino 
oscillations\cite{SuperK98a} and the evidence for
solar neutrino deficits\cite{solarEx} in comparison
to the Standard Solar Model\cite{Bahcall} demonstrate
that the neutrinos are massive. Phenomenological analyses
favor two solutions of the solar and atmospheric neutrino problems,
see e.g.\cite{FLMS}--\cite{BW99a} and for a recent review 
see \cite{BiGG}:
\begin{itemize}
\item 
large $\nu_\mu -\nu_\tau$ mixing for atmospheric anomalies and
matter enhanced\cite{MSW} small mixing angle oscillations (SMA MSW) 
for solar neutrinos, see e.g.\cite{SMAMSW}.
\item 
vacuum oscillations and 
bimaximal\cite{BPWW98}--\cite{BHKR}
or nearly bimaximal\cite{nearly-bimax} mixing of three light neutrinos.
\end{itemize}
We believe that one of these two possibilities is true although other
options are not completely excluded, see e.g.\cite{BKS,Langacker,BiGG,Str99}.
In a near future experimental information from SuperKamiokande
and SNO will select the best scenario. In fact a recent measurement of 
day--night asymmetries\cite{day-night} 
imposes constraints on SMA MSW solutions. On the other hand the vacuum
oscillation solutions may be confirmed or excluded by final results
on seasonal effects
and recoil energy distributions, see \cite{BFLi,BW99a,GKK}
and references therein. A preliminary data from 
SuperKamiokande\cite{prelSK}
are analyzed in Ref.\cite{BW99a}. 

In this article we discuss some consequences
of the bimaximal mixing scenario.
Evidently solutions of `just so' type\cite{just-so} are preferred by the 
existing data. However we do not exclude a possibility\cite{BGG}
that a larger range of $\Delta m^2_\odot$ is allowed for solar neutrinos
if errors are underestimated in one of the radiochemical measurements.
In such a case `just so' fits are also better\cite{BFLi,BW99a}. 
In the following
discussion we use as input parameters\cite{Suzuki99}
$$\Delta m^2_\odot = 4.3\times 10^{-10}\ \mbox{eV}^2$$
for oscillations of solar and
$$ 1.5\times 10^{-3}\ \mbox{eV}^2
\le  \Delta m^2_{atm} \le 6\times 10^{-3}\ \mbox{eV}^2$$
for oscillations of atmospheric neutrinos. 

\section{Maki--Nakagawa--Sakata lepton mixing matrix}
The Dirac masses of quarks and charged leptons exhibit a strongly ordered
hierarchical structure
$$ m_u \ll m_c \ll m_t\  , \quad m_d \ll m_s \ll m_b\ , \quad
m_e \ll m_\mu \ll m_\tau \ .$$
We assume a similar hierarchical structure for the Dirac masses of neutrinos.
However this hierarchical structure is drastically modified by huge Majorana
masses of righthanded neutrinos. As a result of the seesaw 
mechanism\cite{seesaw} three
light and nearly lefthanded neutrinos appear in the mass spectrum.
Their Majorana masses can form patterns very different from the hierarchical 
one. For the same reason the structures of quark and lepton mixing 
matrices are also
quite different. For quarks the Cabibbo-Kobayashi-Maskawa  matrix\cite{CKM} 
$V_{CKM}$ is nearly diagonal and its largest off-diagonal elements
\begin{equation}
| V_{us}|  \simeq | V_{cd} | \approx \theta_c \approx 0.22
\end{equation}
are fairly small. All other off-diagonal elements of $V_{CKM}$ are much
smaller and can be parameterized by higher powers of the Cabibbo angle 
$\theta_c$~\cite{WolfCKM}. $V_{CKM}$ can be written as a product of two 
matrices describing unitary transformations of quarks with weak isospin 
projections $I_3 = + 1/2$ ($u,c,t$) and $I_3 = - 1/2$ ($d,s,b$) respectively:
\begin{equation}
V_{CKM} = {V_{+}}^\dagger\, V_{-} \ \ .
\end{equation}
In the Standard Model only $V_{CKM}$ is observable. In models of quark 
masses and mixing both matrices $V_{\pm}$ are specified and related to
Yukawa couplings of up and down type quarks. A particularly interesting
approach\cite{Fritzsch} relates angles in $V_{+}$ and $V_{-}$ to
mass ratios of up and down type quarks respectively. In this way a relation 
is derived for the mixing between the first two generations
\begin{equation}
\theta_c = \left| 
\sqrt{{m_d/ m_s}} + e^{i\phi} 
\sqrt{{m_u/ m_c}} \right|
\approx \sqrt{{m_d/ m_s}}\ \ .
\end{equation}
The mixing angle $\theta_c$ is dominated by the contribution from the 
$I_3 = - 1/2$ sector.

The flavor mixing matrix\cite{MNS} for three light neutrinos relates
the neutrino mass and flavor eigenstates:
\begin{equation}
\left( \matrix{ \nu_e \cr \nu_\mu \cr  \nu_\tau \cr}\right) =
\left( \matrix{ U_{e1}   &  U_{e2}    &  U_{e3}     \cr
                U_{\mu 1} &  U_{\mu 2}  &  U_{\mu 3}   \cr
                U_{\tau 1}&  U_{\tau 2} &  U_{\tau 3}  \cr } \right)
\left(\matrix{ \nu_1 \cr \nu_2 \cr  \nu_3 \cr}\right) =   
U_{MNS} \left(\matrix{ \nu_1 \cr \nu_2 \cr  \nu_3 \cr}\right) \ \ .    
\label{eq:MNS}
\end{equation}
$U_{MNS}$ can be written as a product of two matrices  $U_{\pm}$
describing transformations of the neutrinos ($I_3 = +1/2$) and the 
charged leptons ($I_3 = -1/2$):
\begin{equation}
U_{MNS} = {U_{-}}^\dagger\, U_{+}\quad .
\label{eq:MNS1}
\end{equation}
It is plausible that the pattern of mixing angles in $U_{-}$
describing the charged lepton sector resembles that in $V_{-}$ 
for the down type quarks, i.e. there is an appreciable mixing in
the electron--muon plane whereas the other mixing angles are 
much smaller
\begin{equation}
U_{-} \approx \left(\matrix{  c  &   -s   &     0   \cr
                                 s  &    c   &     0   \cr
                                 0  &    0   &     1   \cr}\right)
\label{eq:U-1/2}
\end{equation}
where\cite{Fritzsch} 
\begin{equation}
|s| \simeq \sqrt{m_e/m_\mu} \simeq \theta_c/3 \simeq 0.07 \qquad 
{\rm and} \qquad
 c = \sqrt{ 1 - s^2} \approx 1 - s^2/2 \ \ .
\label{eq:s}
\end{equation}
The second equality in (\ref{eq:s}) follows from grand unification
relations\cite{GJ79}
\begin{equation}
m_d \simeq  3m_e\ \ ,     \qquad      m_s \simeq m_\mu/3  \ \ .
\end{equation}
In the sector $I_3 = + 1/2$ the situation is quite different. The structure
of the matrix $U_{+}$ is not directly related to the Dirac mass matrix
$\bf m_{_D}$ for neutrinos. It is strongly affected by the structure of the
Majorana mass matrix $\bf M_{_R}$ for the righthanded neutrinos.
As a consequence the structure of $U_{+}$ is different from the
structure of $V_{+}$. We assume that the mixing of solar and atmospheric
neutrinos is close to bimaximal,
\begin{equation}
U_{+} \approx U_{bm} =
\left(\matrix{   1/\sqrt{2}   &   1/\sqrt{2}   &   0           \cr
                 - 1/2       &     1/2       &   1/\sqrt{2}   \cr 
                   1/2       &    - 1/2      &   1/\sqrt{2}   \cr}\right)\ \ .
\label{eq:U+1/2}
\end{equation}
We also assume that the dominant contribution
to deviations from the ideal bimaximal mixing originates from the 
sector $I_3 = - 1/2$, i.e. from the form of the matrix $U_{-}$.
(Later in this paper we will demonstrate simple seesaw models 
in which corrections to eq.(\ref{eq:U+1/2}) are fairly small.)
The main goal of the present paper is to describe phenomenological
consequences of these two assumptions. For the sake of simplicity we 
neglect CP violation and consider all elements of 
$U_{MNS}$ as real. Then the lepton mixing matrix is completely specified.
From eqs.(\ref{eq:MNS1}), (\ref{eq:U-1/2}) and (\ref{eq:U+1/2}) we obtain
\begin{equation}
U_{MNS} \approx 
\left( \matrix{  
c/\sqrt{2} -s/2\ ,  &   c/\sqrt{2} + s/2\ ,    &    s/\sqrt{2} \cr
-c/2 - s/\sqrt{2}\ ,&   c/2 - s/\sqrt{2}\ ,    &    c/\sqrt{2} \cr
1/2                 &      - 1/2               &    1/\sqrt{2} \cr}  
\right)
\label{eq:MNS2}  
\end{equation}
with $s$ and $c$ given in eq.(\ref{eq:s}).

\section{Neutrino oscillations}
It is well known\cite{BiG98,BW98a} that for 
$\Delta m^2_\odot \ll \Delta m^2_{atm}$ and small $|U_{e3}|$,
oscillations of solar neutrinos effectively decouple from
oscillations in long baseline (LBL) experiments and those of
atmospheric neutrinos. Both types of oscillations can be described
by effective two--neutrino oscillations characterized by phase
differences
\begin{equation}
\Delta_\odot = { \Delta m^2_\odot L\over 4E}
\label{eq:Delsun}
\end{equation}
for solar, and
\begin{equation}
\Delta_{atm} = { \Delta m^2_{atm} L\over 4E}
\label{eq:Delatm}
\end{equation}
for LBL and atmospheric oscillations. In the above equations $L$ 
is the distance traveled by a neutrino of energy $E$.
Probabilities of transitions between different neutrino flavors
in atmospheric and LBL experiments are given by, see \cite{BiGG}
and references therein:
\begin{eqnarray}
{\cal P}_{LBL}\left(\nu_\alpha\rightleftharpoons\nu_\beta\right)
& = & A_{\alpha\beta} \sin^2 \Delta_{atm} \qquad (\alpha\ne\beta) 
\nonumber\\
{\cal P}_{LBL}\left(\nu_\alpha \rightarrow \nu_\alpha \right)
& = & 1 - B_\alpha \sin^2\Delta_{atm} \ \ ,
\end{eqnarray}
where ($\alpha,\beta = e, \mu, \tau$)
\begin{eqnarray}
A_{\alpha\beta} &=&  4 | U_{\alpha3}|^2 | U_{\beta3}|^2 \ \ ,\nonumber\\
B_\alpha        &=&  4 | U_{\alpha3}|^2 \left( 1 - | U_{\alpha3}|^2
\right) \ \ .
\end{eqnarray}
Then eq.(\ref{eq:MNS2}) implies
\begin{equation}
\begin{tabular}{lll}
$A_{e\mu} = s^2 c^2$ ,  \qquad & $ A_{e\tau} = s^2$ , \qquad  & $A_{\mu\tau}= c^2$ ,\\
$B_e = 1 - c^4$ , \qquad & $ B_\mu  = 1 -s^4$ , \qquad  & $ B_\tau = 1$ ,\\
\end{tabular}
\end{equation}
and using (\ref{eq:s}) we estimate that
for appearance type $\nu_e \rightleftharpoons \nu_\mu$ and
$\nu_e \rightleftharpoons \nu_\tau$ LBL experiments
\begin{equation}
\sin^2 2\vartheta_{a} \approx 0.005
\label{eq:appear}
\end{equation}
whereas
\begin{equation}
\sin^2 2\vartheta_{d} \approx 0.01
\label{eq:disappear}
\end{equation}
for disappearance of $\nu_e$ and $\bar\nu_e$ neutrinos. These
predictions can be tested in future high precision experiments.
However the numbers are rather small. The estimation 
(\ref{eq:appear}) is about two times smaller than planned 
sensitivity of MINOS\cite{MINOS}
and eq.(\ref{eq:disappear}) predicts
disappearance of electron neutrinos at the level 20 times
smaller than the present limit from CHOOZ\cite{CHOOZ}. Still
these estimations are larger than those following from
SMA MSW solutions for which $\sin^2 2\vartheta_{d}$
is driven to much smaller values by the recent data on day-night
asymmetries for solar neutrinos\cite{day-night}.

For solar neutrinos, c.f.\cite{BiGG} and references therein,
\begin{equation}
{\cal P}_\odot \left(\nu_e\rightarrow\nu_e\right) =
\left( 1 - | U_{e3} |^2 \right) 
\left( 1 - \sin^2 2\vartheta_\odot\; \sin^2\Delta_\odot \right)
+ | U_{e3} |^4\ \ , 
\end{equation}
where
\begin{equation}
\sin^2 2\vartheta_\odot = 4 | U_{e1} |^2\, | U_{e2} |^2\, / \,
\left( 1 - | U_{e3} |^2 \right)^2 \ \ .
\end{equation}
For the mixing matrix (\ref{eq:MNS2})
\begin{equation}
{\cal P}_\odot = \left( 1- s^2\right)
\left( 1 - \sin^2 2\vartheta_\odot\; \sin^2\Delta_\odot \right) 
+ {\cal O}(s^4)\ \ ,
\end{equation}
with
\begin{equation}
\sin^2 2\vartheta_\odot \approx 1 - s^2 \approx 0.99 \quad . 
\label{eq:thetasun}
\end{equation}
It is evident that the simple picture proposed in this paper 
can be falsified when $\sin^2 2\vartheta_\odot$ derived
from solar neutrino data is significantly below 1. In Ref.\cite{BKS}
the best fit to vacuum oscillations is
\begin{equation}
\Delta m^2_\odot = 6.5 \times 10^{-11}\ \mbox{eV}^2, \qquad 
\sin^2 2\vartheta_\odot = 0.75 
\label{eq:BKS}
\end{equation}
which seems to be in conflict with the estimation in
(\ref{eq:thetasun}). 
In Ref.\cite{BW99a}, see also \cite{BFLi},
a new preliminary data
from SuperKamiokande \cite {Suzuki99} are included into analysis
leading to acceptable solutions in three regions (called $A$, $C$ 
and $D$) of parameter space
\begin{equation}
\matrix{
C: \qquad  & \Delta m^2_\odot = 4.4 \times 10^{-10}\ \mbox{eV}^2, \qquad  &
          \sin^2 2\vartheta_\odot = 0.93  , \qquad  &   gof = 14\%   \cr
D: \qquad  & \Delta m^2_\odot = 6.4 \times 10^{-10}\ \mbox{eV}^2, \qquad  &
          \sin^2 2\vartheta_\odot = 1.00  , \qquad  &   gof = 8\%   \cr
A: \qquad  & \Delta m^2_\odot = 6.5 \times 10^{-11}\ \mbox{eV}^2, \qquad  &
          \sin^2 2\vartheta_\odot = 0.70  , \qquad  &   gof = 6\%   \cr}
\label{eq:BW99a}
\end{equation}
In eq.(\ref{eq:BW99a}) the parameter $gof$ (goodness-of-fit) is the
probability that a random repeat of the given experiment would observe
a greater $\chi^2$, assuming the model is correct. The region $A$ is
the same as selected by the fits of Ref.\cite{BKS}. 
Clearly the analysis
of Ref.\cite{BW99a} gives results in better agreement with 
eq.(\ref{eq:thetasun}). 
It is also seen that new data on solar neutrinos
will  seriously test the estimations presented in this article.  

\section{Neutrinoless double beta decay}
Let us discuss now predictions for neutrinoless double beta decays
($0\nu2\beta$)\footnote{A model independent analysis is presented
in a recent preprint\cite{BW99b}.}. 
Probabilities of such transitions depend on a mass
parameter $B$ which is equal to the absolute value 
of the element $\left( N_A \right)_{11}$ of the matrix
\begin{equation}
N_A  = U_{MNS}\, M_{L,A}\, {U_{MNS}}^T  \quad . 
\end{equation}
$M_{L,A}$ (for $A = I, II, III, IV$) denotes a diagonal matrix of 
Majorana masses for the three light neutrinos:
\begin{equation}
M_{L,A} = m\; diag(\lambda_1,\lambda_2,\lambda_3) \quad  ,
\end{equation}
where $|\lambda_i| \le 1 + {\cal O}(\epsilon,\xi)$;  $\epsilon$ and
$\xi$ are small parameters to be discussed in the following section.
In our convention the eigenvalues $\lambda_i$ are ordered in such a way
that
\begin{equation}
\Delta m^2_\odot/ m^2  = | \lambda_1^2 - \lambda_2^2 |
\end{equation}
and
\begin{equation}
\Delta m^2_{atm}/ m^2  \approx | \lambda_1^2 - \lambda_3^2 |
                       \approx | \lambda_2^2 - \lambda_3^2 | \quad .
\end{equation}
For bimaximal mixing $\nu_e$ is an equal mixture of the mass eigenstates
1 and 2.

Considering $\Delta m^2_\odot /\Delta m^2_{atm}$ as a small perturbation
one obtains $|\lambda_1| = |\lambda_2|$. Then,
up to irrelevant sign redefinitions there are only four
mass patterns consistent with the data on $0\nu 2\beta$:
\begin{enumerate}
\item[I)]
$\qquad m = \sqrt{\Delta m^2_{atm}}$\ , \quad
$\lambda_1 = \lambda_2 = 0$, $\lambda_3 = 1$
\item[II)]
$\qquad  m = \sqrt{\Delta m^2_{atm}}$\ , \quad
$\lambda_1 = -1$, $\lambda_2 = 1$, $\lambda_3 = 0$
\item[III)]
$\qquad  m = \sqrt{\Delta m^2_{atm}}$\ , \quad
$\lambda_1 = \lambda_2 = 1$, $\lambda_3 = 0$
\item[IV)]
$\qquad   m = {\cal O}(1\ \mbox{eV}) \gg \sqrt{\Delta m^2_{atm}}$\ , \quad
$\lambda_1 = -1$, 
$\lambda_2 = 1$, $\lambda_3 = \eta$ where $\eta = \pm 1$ .
\end{enumerate}
In case $IV$ $\Delta m^2_{atm}/m^2$ is considered as a small 
perturbation\footnote{The case $\Delta m^2_{atm}/m^2 ={\cal O}(1)$
requires much more tuning of parameters than the case $IV$. We do not
discuss this case in the present article because it is not clear if a 
reasonable model of this kind exists. Furthermore an original motivation
for degenerate neutrino masses is an appreciable neutrino contribution
to dark matter in the Universe whereas this contribution is small if 
$\Delta m^2_{atm}/m^2$ is not small.}.
The states 1 and 2 have opposite CP parities (i.e. 
$\lambda_1$ and $\lambda_2$ have opposite signs, see e.g.\cite{BiGG})
due to the experimental bound\cite{Baudis}:
\begin{equation}
B < 0.2\ \mbox{eV} 
\end{equation}
which is violated for $m > 0.2\ eV$ and $\lambda_1 = \lambda_2 = 1$. 
The argument is analogous to the case $III$ which is considered in
the following, c.f. eq.(\ref{eq:NIII11}). 
\par\noindent
If corrections to bimaximal mixing are neglected the following mass matrices
are obtained
\begin{eqnarray}
{\bar N_I} &=& {\textstyle {1\over2} } \sqrt{\Delta m^2_{atm}}
\left( 
\matrix{0  &   0    &   0  \cr
        0  &   1    &   1  \cr
        0  &   1    &   1  \cr } \right) \quad , 
\label{eq:textureI}\\
{\bar N_{II}} &=& {\textstyle {1\over\sqrt{2}}} \sqrt{\Delta m^2_{atm}}
\left(
\matrix{\ 0  &   1    &  -1  \cr
        \ 1  &   0    &  \ 0  \cr
        -1  &   0    &  \ 0  \cr } \right) \quad , 
\label{eq:textureII}\\
{\bar N_{III}} &=& {\textstyle {1\over2} } \sqrt{\Delta m^2_{atm}}
\left(
\matrix{2  &  \ 0    & \  0  \cr
        0  &  \ 1    &   -1  \cr
        0  &  -1     & \  1  \cr } \right) \quad , 
\label{eq:textureIII}\\
{\bar N_{IV} } &=& {\textstyle {1\over2} }\,  m \,
\left(
\matrix{\ 0        & \sqrt{2}  &   -\sqrt{2}  \cr
     \ \sqrt{2}    & \eta      &   \ \eta      \cr
      -\sqrt{2}    & \eta      &   \ \eta      \cr} \right)\quad ,
\label{eq:textureIV}
\end{eqnarray}
where for $ A = I, II, III, IV$:
\begin{equation}
\bar N_{A} = U_{bm}\; M_{L,A} \; {U_{bm}}^T  \quad  .
\end{equation}
In the following section we consider also matrices
\begin{equation}
\widetilde N_{A} = U_{+}\; M_{L,A} \; {U_{+}}^T  
\end{equation}
which include deviations from the ideal bimaximal mixing due to the sector
$I_3 = + 1/2$. The textures (\ref{eq:textureI})--(\ref{eq:textureIV}) 
have been discussed in the
literature\cite{JS98}--\cite{FGN}. 
It is seen that only
$\bar N_{III}$ leads to non-zero rates of $0\nu2\beta$ decays. In this case
deviations from bimaximal mixing can be neglected and one obtains the 
following prediction
\begin{equation}
| N_{III} |_{11} \approx  \sqrt{\Delta m^2_{atm}}
= ( 6 \pm 2 ) \times 10^{-2}\ \mbox{eV} \quad .
\label{eq:NIII11}
\end{equation}
In the three other cases the role of $U_{-}$ is essential. Keeping only
leading terms one obtains
\begin{eqnarray}
| N_{I} |_{11} \approx& {\textstyle{1\over2}}\; s^2\;
\sqrt{\Delta m^2_{atm}}
&= ( 1.5 \pm 0.5 ) \times 10^{-4}\ \mbox{eV} \quad ,
\label{eq:NI11}\\
| N_{II} |_{11} \approx& \sqrt{2}\, |s|\, 
\sqrt{\Delta m^2_{atm}}
&= ( 6 \pm 2 ) \times 10^{-3}\ \mbox{eV} \quad ,
\label{eq:NII11}\\
| N_{IV} |_{11} \approx& \sqrt{2}\, |s|\, m &\approx 0.1\, m \quad .
\label{eq:NIV11}
\end{eqnarray}
Evidently the cases $III$ and $IV$ can be confirmed or ruled out by
next generation experiments which may be sensitive to $B$ as low as
0.01 eV\cite{K-KHH}. Even the present limit\cite{Baudis} implies that
for degenerate masses $m$ must be smaller than 2~eV. 

\section{Patterns of neutrino masses}
In this section effects are considered due to non-zero value of the 
ratio $\Delta m^2_\odot /\Delta m^2_{atm}$ which were neglected in the
discussion of the preceding section. We present simple seesaw models 
which show that these corrections are quite small, so the estimations
(\ref{eq:NIII11})--(\ref{eq:NIV11}) are not much affected. 
In the following we consider 
the four cases $I$--$IV$ corresponding to different neutrino mass patterns.
\subsection{Case I} A class of seesaw models corresponding to case $I$
has been described in\cite{JS98}. In these models the third mass eigenstate
is much heavier than the other two whose masses are fairly close. 
In the following this mass pattern ($m_1\approx m_2 \ll m_3$) 
is called {\em semi-hierarchical}. In terms of a small parameter
\begin{equation}
|\epsilon| = \left( {\Delta m^2_\odot\over 2\Delta m^2_{atm}} \right)^{1/3}
\approx 0.004
\end{equation}
the eigenvalues $\lambda_i$ are given by the expansions:
\begin{eqnarray}
\lambda_1 &=& \ \epsilon - \epsilon^2/2 + \dots  \nonumber\\
\lambda_2 &=& -\epsilon - \epsilon^2/2 + \dots  \nonumber\\
\lambda_3 &=& 1 + \epsilon^2 + \dots 
\label{eq:semi-hierarch}
\end{eqnarray}
For semi-hierarchical mass pattern $\nu_e$ is a linear combination of the 
two lighter mass eigenstates of opposite CP parities. The element
$(\widetilde N_{I} )_{11}$ of the matrix
\begin{equation}
\widetilde N_{I} = U_{+}\; M_{L,I} \; {U_{+}}^T 
\end{equation}
is strongly suppressed and the main contribution to $0\nu 2\beta$
transitions originates from the mixing in the sector $I_3 = - 1/2$.
The estimation of $| N_{I} |_{11}$ given in eq.(\ref{eq:NI11})
is not affected as can be seen from the explicit form of the matrix
$U_{+}:$\footnote{ In notation of Ref.\cite{JS98}  
$r = 2^{3/2}\epsilon$. The columns of the matrix $U_{+}$ in 
eq.(\ref{eq:U+1/2a}) are normalized eigenvectors $v_1$, $- v_2$ and $v_3$,
c.f. eq.(29) in \cite{JS98}. We have corrected a misprint in the second
element of the vector $v_2$ in \cite{JS98}.}
\begin{equation}
U_{+} = \left( \matrix{  
( 1 + \epsilon/2 )/\sqrt{2}    &   ( 1 - \epsilon/2 )/\sqrt{2}     &   \epsilon\cr
 - 1/2 -  \epsilon             &   1/2 -  \epsilon                 &    1/\sqrt{2} \cr     
        1/2                    &          - 1/2                    &    1/\sqrt{2} \cr }   
\right) + {\cal O}(\epsilon^2)   =
U_{bm} +  {\cal O}(\epsilon)
\label{eq:U+1/2a}
\end{equation}
and
\begin{equation}
\widetilde N_I = {\textstyle {1\over 2}} \left( \matrix{
0               &      0      &  2^{3/2}\epsilon  \cr
0               &      1      &         1         \cr
2^{3/2}\epsilon &      1      &         1         \cr  } \right)
+  {\cal O}(\epsilon^2) \quad .
\label{eq:tildeNI}
\end{equation}

In case $I$ a strongly ordered {\em hierarchical} mass pattern 
($m_1 \ll m_2 \ll m_3$) is  obtained if e.g.
\begin{equation}
\lambda_1 = \epsilon^2 + \dots \ , \qquad  \lambda_2 = \epsilon + \dots \ ,
\qquad \lambda_2 = 1 + \dots \ ,
\end{equation}
with 
\begin{equation}
 |\epsilon| = \sqrt{\Delta m^2_\odot /\Delta m^2_{atm}} 
\approx 3.7 \times 10^{-4} \ \ . 
\end{equation}
The magnitude of $\epsilon$ is an order of magnitude smaller than
for semi-hierarchical spectrum. There is no suppression of 
$\left( N_{I} \right)_{11}$ because the lighter states have the
same CP parity but the parameter $\epsilon$ is small.
Seesaw models leading to hierarchical 
spectrum may require some fine tuning. Thus we believe that 
hierarchical mass pattern is less attractive from theoretical point
of view. In spite of theoretical prejudices we note that for
both patterns predictions for oscillations and $0\nu 2\beta$ decays
are similar.
\subsection{Case II} Corresponding seesaw models can be easily derived.
An example is (in notation of \cite{JS98}):
\begin{equation}
{\bf m_{_D}}  \sim \left( \matrix{
x^2 y     &      0      &     0    \cr
0         &      x      &    -x    \cr
0         &      x^2    &     1    \cr}  \right)
\qquad   {\rm and} \qquad  
{\bf M_{_R}^{-1}}  \sim  \left( \matrix{
a_{11}    &      a_{12}     &     0   \cr
a_{12}    &         0       &     0   \cr
  0       &         0       &     0   \cr} \right)\quad  ,
\end{equation}
where ${\bf m_{_D}}$ is the Dirac mass matrix and ${\bf M_{_R}^{-1}}$
denotes inverse of the Majorana mass matrix for the righthanded
neutrinos\footnote{
This form of ${\bf M_{_R}^{-1}}$ means that either one of the righthanded
neutrinos is much heavier and decoupled or there are only two
heavy neutrinos in the particle spectrum
and their couplings to the lefthanded neutrinos are given by the matrix
$$ {\bf m_{_D}} \sim \left( \matrix{  
xy   &     0      &     0    \cr
0    &     1      &    -1    \cr} \right)  .$$ }. Then 
\begin{equation}
\widetilde N_{II} = U_{+}\; M_{L,II} \; {U_{+}}^T =
{\bf m_{_D}}^T\, {\bf M_{_R}^{-1}}\, {\bf m_{_D}} =
\sqrt{\Delta m^2_{atm}}\, \left( \matrix{
2 \epsilon    & 1/\sqrt{2}       &     -1/\sqrt{2} \cr
1/\sqrt{2}    &       0          &          0      \cr
 -1/\sqrt{2}  &       0          &          0      \cr} \right)\quad ,  
\end{equation}
with the eigenvalues
\begin{equation}
 \lambda_1  = -\sqrt{1 + \epsilon^2} + \epsilon\ , \qquad
   \lambda_2  = \ \sqrt{1 + \epsilon^2} + \epsilon\ , \qquad
   \lambda_3 =0 
\end{equation}
and
\begin{equation}
 |\epsilon| = 2^{-3/2} xy |a_{11}/a_{12}| = 
\Delta m^2_\odot / (4 \Delta m^2_{atm}) = 3.4 \times 10^{-8}\quad  .
\end{equation}
Evidently corrections of order $\epsilon$ to $U_{MNS}$ are negligibly
small. The spectrum of light neutrinos contains two heavier states.
The electron neutrino is a nearly maximal mixture of these states
with a tiny admixture of the third state which is massless. We call
this spectrum {\em semi-degenerate} ($m_1 \approx m_2 \gg m_3$).
The ratio $|a_{11}/a_{12}|$ must be very small which implies that the two
heavy Majorana neutrinos form a pseudo-Dirac system\cite{BHS98}.
In our opinion the semi-degenerate mass pattern is more attractive
for solutions of the solar neutrino problem with larger values of
$\Delta m^2_\odot / \Delta m^2_{atm}$~\cite{BGG}, 
see also discussion in \cite{FGN}.
\subsection{Case III} As in case $II$ the parameter $\epsilon$ is tiny
and the related corrections are negligible.
\subsection{Case IV} For {\em degenerate} neutrino masses\cite{Vissani97,GG98} 
 one can choose the 
eigenvalues $\lambda_i$ in the following way:
\begin{equation}
 \lambda_1 = -1 + \epsilon\ , \qquad \lambda_2 = 1 + \epsilon\ , \qquad 
\lambda_3 = \eta + \xi \ ,
\end{equation}
with
\begin{equation}
 |\epsilon| = \Delta m^2_\odot / (4 m^2) \qquad
{\rm and} \qquad
|\xi| = \Delta m^2_{atm} /(2 m^2) \quad .
\end{equation}
Corrections to eq.(\ref{eq:NIV11}) are small for $|\xi| \ll |s|$.

If the case of degenerate neutrino masses
is a reasonable option for
vacuum oscillations remains an open question. On one side dynamical
models have been recently found\cite{BHKR} for the degenerate mass
pattern. On the other side the parameter $\epsilon$ is extremely
small and the problem of stability 
may be very serious, see \cite{ELola}. It is also not clear if neutrino
masses in the range of 1~eV are really needed by cosmology, see e.g.
\cite{Turner}. Fortunately it is exactly this mass range and mass
pattern for which our estimations give numbers close to the present
experimental limit for $0\nu 2\beta$ decays. Therefore one may hope
that a definite answer will come from the experiment in not very
distant future.
\section{Summary}
Neutrino oscillations and Majorana masses are discussed assuming
bimaximal mixing of leptons. In this scheme the atmospheric neutrino
anomaly~\cite{SuperK98a} is described as a result of $\nu_\mu\to \nu_\tau$
oscillations and deficits of solar neutrinos~\cite{solarEx} as vacuum
oscillations of $\nu_e$. A simple hypothesis is proposed that deviations
from bimaximal mixing are dominated by a rotation in the electron--muon
plane of the charged lepton sector $I_3 = - 1/2$. This hypothesis is in
agreement with the present data and can be tested by future precision data
on solar neutrinos. Predictions for oscillations in long baseline 
experiments are given. There are four classes of neutrino mass patterns 
leading to quite different predictions for the rates of neutrinoless
double beta decays. 
\vskip0.5cm\par\noindent
{\Large\bf Acknowledgements}
\vskip0.5cm\par\noindent
M.J. thanks Wolfgang Hollik, Hans K\"uhn, Thomas Mannel and Marek 
Zra{\l}ek for useful comments and remarks. Y.S. is grateful to 
K. Inoue for discussions.

M.J. would like to acknowledge a support from BMBF (FRG) under 
Project Nr. POL-239-96, and a very
stimulating atmosphere and warm hospitality 
during his stay in the Institut f\"ur Theoretische Teilchenphysik,
Universit\"at Karlsruhe where a large part of this work was done.

\newpage

\end{document}